# Minimizing Cache Timing Attack Using Dynamic Cache Flushing (DCF) Algorithm

Jalpa Bani
Computer Science and Engineering Department
University of Bridgeport
Bridgeport, CT 06601
jbani@bridgeport.edu

Syed S. Rizvi
Computer Science and Engineering Department
University of Bridgeport
Bridgeport, CT 06601
srizvi@bridgeport.edu

*Abstract*—**Rijndael algorithm was unanimously chosen as the Advanced Encryption Standard (AES) by the panel of researchers at National Institute of Standards and Technology (NIST) in October 2000. Since then, Rijndael was destined to be used massively in various software as well as hardware entities for encrypting data. However, a few years back, Daniel Bernstein [2] devised a cache-timing attack that was capable enough to break Rijndael's seal that encapsulates the encryption key. In this paper, we propose a new Dynamic Cache Flushing (DCF) algorithm which shows a set of pragmatic software measures that would make Rijndael impregnable to cache timing attack. The simulation results demonstrate that the proposed DCF algorithm provides better security by encrypting key at a constant time.**

Keywords- dynamic cache flushing, Rijndael algorithm, timing attack.

I. INTRODUCTION

Rijndael is a block cipher adopted as an encryption standard by the U.S. government. It has been analyzed extensively and is now used widely worldwide as was the case with its predecessor, the Data Encryption Standard (DES). Rijndael, the AES standard is currently used in various fields. Due to its impressive efficiency [8], it's being used in high-speed optical networks, it's used in military applications that encrypt top secret data, and it's used in banking and financial applications wherein secured and real-time transfer of data is a top-priority.

Microsoft has embraced Rijndael and implemented Rijndael in its much talked about DotNet (.NET) Framework. DotNet 3.5 has Rijndael implementation in System.Security.Cryptography namespace. DotNet framework is used by millions of developers around the world to develop software applications in numerous fields. In other words, software implementation of Rijndael is touching almost all the fields that implements cryptography through the DotNet framework.

Wireless Network Security has no exception. Wired Equivalent Privacy (WEP) is the protocol used in wireless networks to ensure secure environment. When WEP is turned on in a wireless network, every packet of data that is transmitted from one station to another is first encrypted using Rijndael algorithm by taking the packets' data payload and a secret encryption key called WEP key. The encrypted data is then broadcasted to stations registered on that wireless network. At the receiving end, the "wireless network aware stations" utilize the WEP key to decrypt data using Rijndael algorithm. Rijndael supports a larger range of block and key sizes; AES has a fixed block size of 128 bits and a key size of 128, 192 or 256 bits, whereas Rijndael can be specified with key and block sizes in any multiple of 32 bits, with a minimum of 128 bits and a maximum of 256 bits [6].

This algorithm implements the input, output, and cipher key where each of the bit sequences may contain 128, 192 or 256 bits with the condition that the input and output sequences have the same length. However, this algorithm provides the basic framework to make the code scalable. Look up tables have been used to make Rijndael algorithm faster and operations are performed on a two dimensional array of bytes called states. State consists of 4 rows of bytes, each of which contains Nb bytes, where Nb is the input sequence length divided by 32. During the start or end phase of an encryption or decryption operation, the bytes of the cipher input or output are copied from or to this state array.

The several operations that are implemented in this algorithm are listed below [9]:

- Key Schedule: It is an array of 32-bit words that is initialized from the cipher key. The cipher iterates through a number of the cycles or rounds, each of which uses Nk words from the key schedule. This is considered as an array of round keys, each containing Nk words.

- Finite Field Operations: In this algorithm finite field operations are carried out, which refers to operations performed in the finite field resulting in an element within that field. Finite field operations such as addition and multiplication, inverse multiplication, multiplications using tables and repeated shifts are performed.

- Rounds: At the start of the cipher the input is copied into the internal state. An initial round key is then added and the state is then transformed by iterating a round function in a number of cycles. On completion the final state is copied into the cipher output [1].

The round function is parameterized using a key schedule that consists of a one dimensional array of 32-bit words for which the lowest 4, 6 or 8 words are initialized with the cipher. There are several steps carried out during this operation:

SubBytes: As shown in Fig. 1, it is a non-linear substitution step where each of the byte replaces with another according to a lookup table.

ShiftRows: This is a transposition step where each row of the state is shifted cyclically a certain number of steps, as shown in Fig. 2.

MixColumns: This is a mixing operation which operates on the columns of the state, combining the four bytes in each column, as shown in Fig. 3.

AddRoundKey: Here each byte of the state is combined with the round key; each round key is derived from the cipher key using a key schedule [1], as shown in Fig. 4.

- Final Round: The final round consists of the same operations as in the Round function except the MixColumns operation.

## II. RELATED WORK

Parallelism or Parallel Computing has become a key aspect of high performance computing today and its fundamental advantages have deeply influenced modern processor designers. It has become a dominant paradigm in processor architecture in form of multicore processors available in personal computers today. Sharing processor resources like cache memory, sharing memory maps in random access memory (RAM) and sharing computational power of the math coprocessors during execution of multiple processes in the operating systems, has become an inevitable phenomenon. Few years back, Intel introduced hyper-threading technology in its Pentium 4 processors, wherein the sharing of processor resources between process threads is extended further by sharing memory caches. Shared access to memory cache is a feature that's available in all the latest processors from Intel and AMD Athlon.

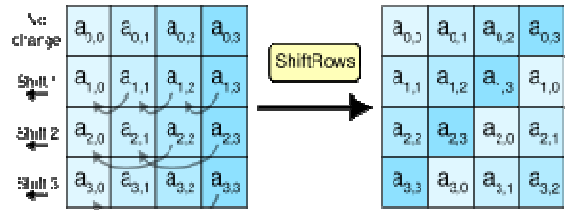

Figure 2. ShiftRows

With all the hunky-dory talk about how parallel computing has made Central Processing Unit's (CPUs) very powerful today, the fundamentals of sharing memory cache across the thread boundary has come along opening doors for security vulnerabilities. The shared memory cache can permit malicious threads of a spy process to monitor execution of another thread that implements Rijndael, allowing attackers to brute force the encryption key [6, 7].

## III. PROBLEM IN RIJNDAEL: CACHE TIMING ATTACK

Cache timing attack – the name speaks for itself. This belongs to a pattern of attacks that concentrates on monitoring the target cryptosystem, and analyzing the time taken to execute various steps in the cryptographic algorithm. In other words, the attack exploits the facts that every step in the algorithm takes a certain time to execute.

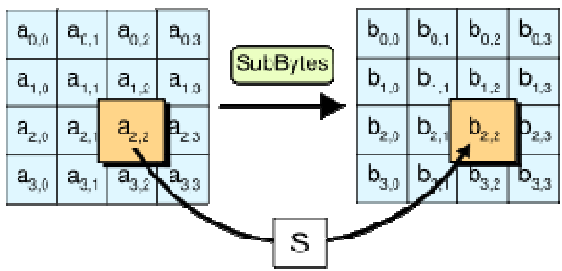

Figure 1. SubBytes

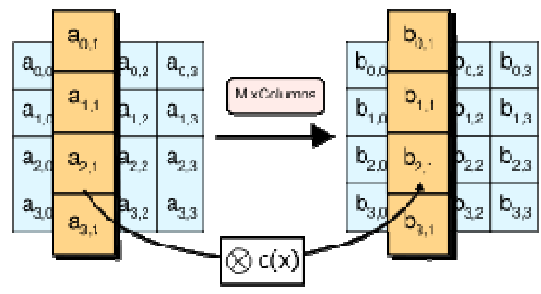

Figure 3. MixColumn

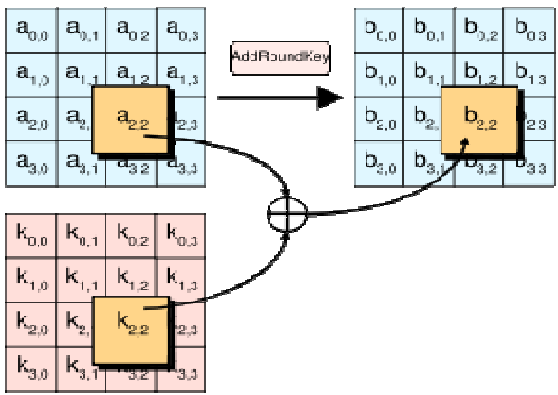

Figure 4. AddRoundKey

Although, the cache-timing attack is well-known theoretically, but it was only until April 2005 that a stout researcher named Daniel Bernstein [2, 4] published that the weakness of Rijndael can reveal timing information that eventually can be utilized to crack the encryption key. In his paper, Daniel announced a successful cache timing attack by exploiting the timing characteristics of the table lookups.

Here is the simplest conceivable timing attack on Rijndael. AES software implementations like Rijndael that uses look-up tables to perform internal operations of the cipher, such as Sboxes, are the one that are most vulnerable to this attack. For example, the variable-index array lookup T0[k[0] $\oplus$ n[0]] near the beginning of the AES computation. A typical hacker might think that the time for this array lookup depends on the array index and the time for the whole AES computation is well correlated with the time for this array lookup. As a result, the AES timings leak information about k[0] $\oplus$ n[0] and it can calculate the exact value of k[0] from the distribution of AES timings as a function of n[0]. Similar comments apply to k[1] $\oplus$ n[1], k[2] $\oplus$ n[2], etc. Assume, that the hacker watches the time taken by the victim to handle many n's and totals the AES times for each possible n[13], and observes that the overall AES time is maximum when n[13] is, say, 147. Suppose that the hacker also observes, by carrying out experiments with known keys k on a computer with the same AES software and the same CPU, that the overall AES time is maximum when k[13] $\oplus$ n[13] is, say, 8. The hacker concludes that the victim's key k[13] is 147$\oplus$ 8 = 155. This implies that a hacker can easily attack a variable time AES algorithm and can crack the encrypted data and eventually key [2].

Since in Rijndael algorithm all look up tables are stored in the cache, by putting another thread or some different way, attacker can easily get the encrypted data from the cache. Fig.1 shows that AES implementation in OpenSSL which does not take constant time. This was taken on a Pentium M processor. It is a 128 x 128 array of blocks where X axis shows one key for each row of blocks and Y axis shows one input for each column of blocks. Any combination of (key, Input) pair shows the encryption process for that particular pair by indicating the fix pattern of colors at that place. We can see the tremendous variability among blocks in Fig. 5. Due to this variability, attacker can easily determine the weak point, where the encryption took place by just analyzing the color pattern.

The cache timing attack problem has been tackled through various approaches [3]. Each solution has its own pros and cons. For instance, Intel released a set of compilers targeting their latest 64-bit processors. These compilers would take the C++ code as input and output a set of machine instructions that would not use CPU cache at all. In other words, the resultant code has a machine instruction that does not use CPU cache for temporary storage of data, in other words the cache is disabled automatically.

The other suggestion was to place all the lookup tables in CPU registers rather than CPU cache, but this would affect performance significantly. Hardware approaches are also being considered. It has been suggested to have a parallel Field-Programmable Gate Array (FPGA) implementation or Application-Specific Integrated Circuits (ASIC) implementation with a separate coprocessor functioning with the existing CPU. This special coprocessor would contain special logical circuitry that would implement Rijndael. Timing attack can thus be avoided by barring other processes from accessing the special coprocessor [5].

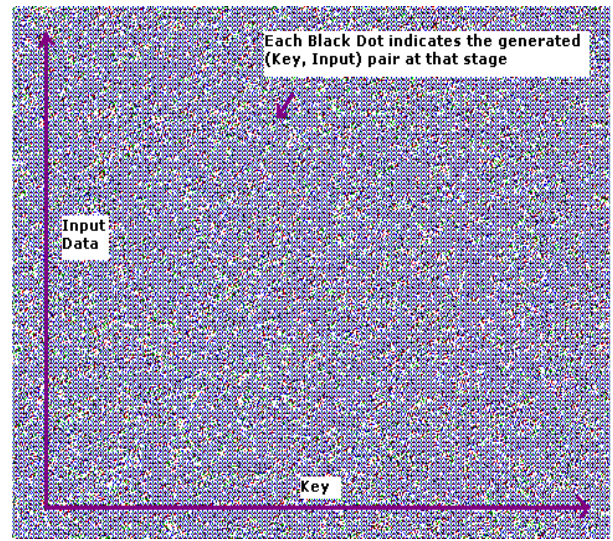

Figure 5. Open SSL AES timings for 128 keys and 128 inputs on a Pentium M processor

## IV. PROPOSED DYNAMIC CACHE FLUSHING (DCF) ALGORITHM

Numerous attempts have been made to address the timing attack loophole in AES. After a deep analysis of the logical steps involved in the Rijndael algorithm, we propose a novel technique to improvise the existing Rijndael algorithm. Our proposed algorithm follows variable-time AES algorithm by replacing it with a constant-time (but not high-speed) AES algorithm known as DCF (Dynamic Cache Flushing). Here, constant means totally independent of the AES key and input. The resulting DCF algorithm would be capable enough to stand strong against the timing attacks.

In order to determine the constant-time, first we need to collect timings and then look for input-dependent patterns. For example, we can repeatedly measure the time taken by AES for once (key; input) pair, convert the distribution of timings into a small block of colors, and then repeat the same color pattern for many keys and inputs.

A constant-time AES algorithm would have the same block of colors for every key and input pair, as shown in Fig 2. Fig 2 is a 128 x 128 array of blocks. Here, X axis indicates the key for each row of blocks and Y axis shows the input for each column of blocks. The pattern of colors in a block reflects the distribution of timings for that (Key; Input) pair. Here, for all (Key, Input) pairs, the color patterns remains the same, due to the constant time. Hence, attacker cannot easily figure out at which point of time the encryption of key and data took place. DCF algorithm generates keys at a constant rate on today's popular dual-core CPUs.

### A. Description of the Proposed DCF Algorithm

The DCF algorithm is the improved version of Rijndael. In other words, the basic encryption/decryption process would remain unchanged. However, there are few additional steps injected into the Rijndael algorithm that would make it resilient to cache-timing attack.

DCF algorithm – as the name rightly suggests, flushes cache while the encryption of data is in progress. In other words, the data that is being copied by the program into the CPU cache during the encryption/decryption process is removed at periodic intervals. The major advantage of doing this is that, during a cache-timing attack, the spy process tries to tap the data stored in look up tables in the CPU cache. Since each instruction takes time to encrypt or decrypt the data, attacker can break the data by just taking difference of collected large body of timing data from the target machine for the plaintext byte and collected large body of reference timing data for each instruction.

Fig. 5 shows that encryption/decryption takes place at random time and it can be easily determined by the spy process. If data in the CPU cache is flushed dynamically during the encryption or decryption process, it would make life more difficult for the spy process, when it tries to collect the data for sampling purposes. In addition, no data in the cache implies that there is no specific place or point that refers to the encryption process as shown in Fig. 6.

It should be noted in Fig. 6 that the graph maintains a uniform pattern during the entire encryption/decryption process. Due to this uniformity, an attacker would face difficulty in tracking the exact time frame when encryption/decryption took place. This is possible by flushing the CPU cache at irregular intervals. Flushing the cache ensures that an attacker will not get enough insight into the data pattern during the encryption process by tapping the cache data. In order to increase the efficiency of this approach, one can increase the frequency of cache flushing. This would be a customizable parameter in the proposed DCF implementation. By further analyzing the DCF algorithm, it would lead to more "cache-misses" than "cache-hits". The "cache-misses" would eventually be recovered by looking up into the RAM for data. The "cache-misses" is the performance penalty we pay with this approach. But with the computing capability we have today with the high-end dual core CPUs, this refetching of data elements from the RAM, can be dealt with.

It should be noted that complete cache disabling is also an option [3], but in such scenarios the spy process might as well start tapping the RAM for encrypted data. Flushing the cache would rather confuse the spy process and make life difficult for attackers to derive a fixed pattern of the timing information and encrypted data samples.

Another feature intended in DCF algorithm is to implement random delays within the execution cycles during the encryption/decryption process. As a matter of fact that if bunch of the instructions from the encryption program repeats more than once, the execution time for those instructions remain constant all the time. By continuously monitoring the CPU instruction cycles, attacker can determine the time taken to execute a step in encryption algorithm. Attacker might be able to capture the entire process timeline and data patterns being encrypted or decrypted. In DCF, additional delays could be introduced while the algorithm steps are in progress. This would change the encryption/decryption timeline and make the algorithm more unpredictable. As a result, attacker will not be able to guess the timing pattern created by the encryption/decryption steps.

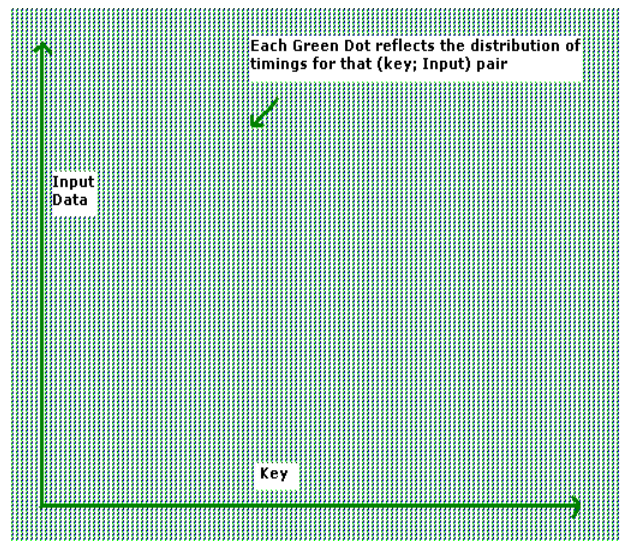

Figure 6. AES timings, using Constant-Time AES algorithm, for 128 keys and 128 inputs

Every time when the proposed DCF algorithm generates a unique timing pattern for encrypting the set of data, it makes things more difficult for an attacker who uses a key parameter (i.e., the time taken to encrypt a set of data) in his predictable brute-force approach for cracking the key. The delays in DCF could be made more unpredictable by randomizing the numeric values that defines the amount of delay caused. A good sturdy randomizer could achieve a fairly unpredictable pattern of Fig. 5 Open SSL AES timings for 128 keys and 128 inputs on a Pentium M processor and Fig. 6 AES timings using Constant-Time AES algorithm, for 128 keys and 128 input delays.

The cache timing attack exploits the effect of memory access on the cache, and would thus be completely lessened by an implementation that does not perform any table lookups. Instead of avoiding table lookup, one could employ them by ensuring that the pattern of accesses to the memory is completely independent of the data passing through the algorithm. In its easiest form, implementing a memory access for a relevant set of data, one can read all the data from the look-up table. In addition, one could use an alternative description of the cipher which replaces the table lookups by an equivalent series of the logical operations. For AES, this is particularly ideal since the lookup tables have concise algebraic descriptions, but performance is degraded by over an order of magnitude [3].

Flushing cache, random delays, and making data access independent of underlying data being processed, would make sense only if the DCF program is forced to run on a single thread. Single thread would also ensure that less data is being exposed to the spy process at any given point of time.

### B. Mathematical Model

As discussed above, Rijndael is vulnerable to timing attacks due to its use of table lookups. In the current analysis, we develop a mathematical model for the attacks when table lookups are being performed during the execution of a Rijndael algorithm. We use our inventive method of flushing the cache during the execution of the table lookups and prove that when the table lookups are performed in constant-time, the attacker is unable to apply his/her spy process to recognize the encrypted data. Fig. 7 and 8 are plotted for constant-time DCF algorithm using a tool called "CacheIn" - a toolset for comprehensive Cache Inspection from Springer. Counter measures like flushing cache are implemented in the DCF algorithm using C++.

Fig. 7 shows the average time taken to execute the ith instruction and $xi0$ indicates the part of the instruction cycle. Here, X axis shows the instruction cycle and Y axis shows the time taken to execute that particular instruction. Fig. 8 shows the average time taken to fetch the input data Pi from the cache for that particular instruction xi0. Here, X axis shows the data in cache memory and Y axis shows the time taken to fetch that data. Due to the constant time approach with the cache flushing, Fig. 7 and Fig. 8 demonstrate that an average time reaches to a constant value. Fig. 9 is the combination of the timing graphs shown in Fig. 7 and 8 for fetching the data and the time taken to execute the instruction to fetch that data.

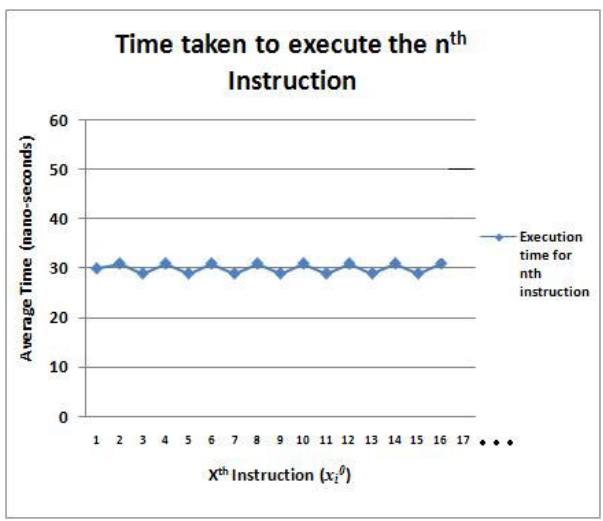

Figure 7. Graph showing time taken to execute the instruction

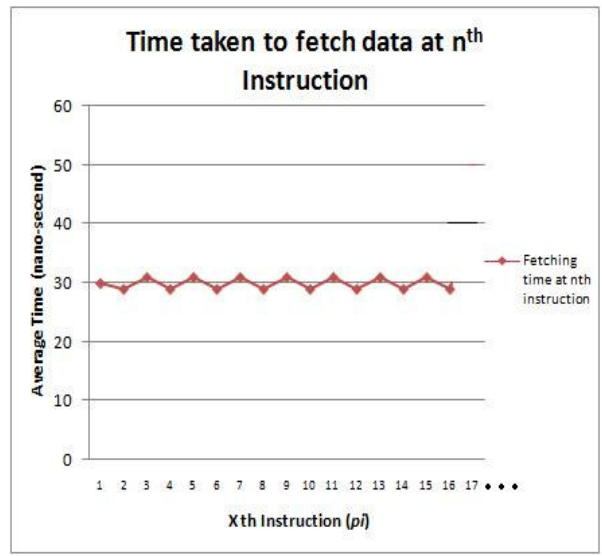

Figure 8. Graph showing time taken to collect data from cache during each CPU instruction.

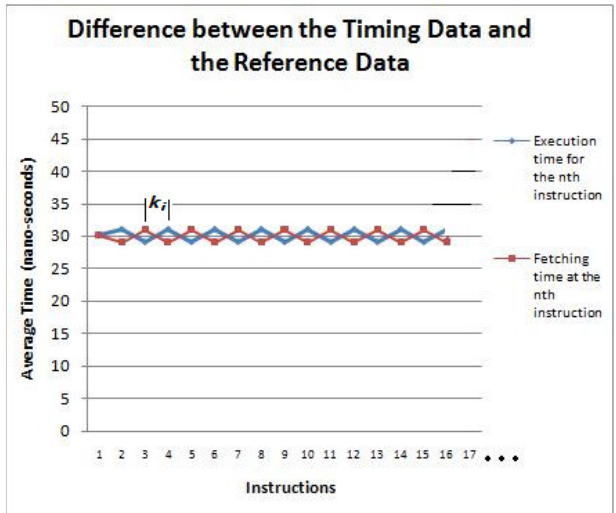

Figure 9. Graph showing the difference between the timing data and reference data

If we take the difference of maximum values of an average time for fetching the data and the time to execute an instruction to fetch that data, we will get very negligible time difference, say ki. For any time difference between the timing data and the reference data, ki remains constant and too small due to cache flushing. This implies that, with the constant time information, it is not possible to determine the exact time taken to encrypt/decrypt the data. The performance of the DCF algorithm is found to be little bit slower than the Rijndael algorithm. The performance penalty is due to cache flushing that provokes the processor to search the missing data in the RAM or in a secondary disk. On the other hand, the security provided against attackers by the proposed DCF algorithm is pretty impressive.

V. SIMULATION RESULTS

Here is a brief description of DCF during execution of Rijndael algorithm. Assume that there is a huge data file that's being encrypted using the DCF algorithm. The flowchart in Fig. 10 would portray a logical flow of events. A huge file is read into a user-defined variable, "buffer". The password provided by the user is typically stored as the encryption key. Rijndael initializes itself by building the set of round tables and table lookups into its data structure which helps in processing the data in buffer. A timer is initialized just before Rijndael starts encrypting the data in the buffer. The time should be initialized in nanoseconds. During encryption, Rijndael puts the key and data together in the round operation. During various steps in the encryption process, the random delays are introduced using *Sleep(X)* function to ensure that the repeated set of instructions does not portray the same execution timeline. Here, the amount of time, the process needs to be suspended 'X', is directly proportional to the total amount of time 'T' taken to process the chunk of data of size 'S'. If the timer becomes zero, flush or remove the data from the cache by using the *cacheflush()* function. The timer would be initialized with a random time that would make the encryption process time more unpredictable for the hacker. Reinitialize the timer with a random time and perform the encryption with random delay until all the data is processed (encrypted).

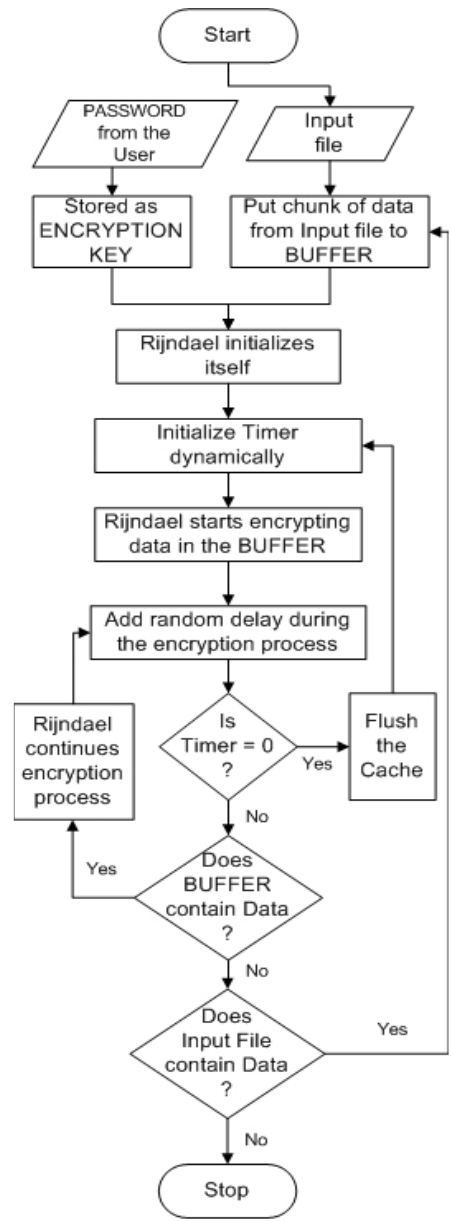

Figure 10. Dynamic Cache Flushing Algorithm Flowchart

## VI. CONCLUSION

We have seen that Rijndael is vulnerable to cache timing attack. Beyond AES, such attacks are potentially applicable to any implementation of a cryptographic algorithm that performs data-dependent memory accesses. The main weakness detected in the Rijndael algorithm is the heavy use of table lookups which dominate the running time and the table lookup indices. The countermeasures described in this paper represent a significant step towards developing a stable, attack-proof AES algorithm. The DCF algorithm simulates a scenario wherein the table lookups are accessed in constant-time rather than in variable-time. This would disable any attacker from writing a spy program to brute force the key and data out of the cache data stored during the execution of the DCF algorithm. In the implementation of the DCF algorithm, cache is flushed periodically during encryption or decryption process. This would disable the attacker from tapping the cache for data. On the downside, there is a performance hit on the encryption time, but on a brighter note, the DCF algorithm stands strong against the cache timing attack.

## REFERENCES

[1]  J. Daemen and V. Rijmen, "AES Proposal: Rijndael, AES Algorithm" Submission, September 3, 1999.

[2]  Daniel J. Bernstein, "Cache-timing attacks on AES", The University of Illinois at Chicago, IL 60607-7045, 2005.

[3]  D.A. Osvik, A. Shamir and E. Tromer. "Cache attacks and Countermeasures: the Case of AES". In Cryptology ePrint Archive, Report 2005/271, 2005.

[4]  Joseph Bonneau and Ilya Mironov, "Cache-Collision Timing Attacks Against AES" , (Extended Version) revised 2005-11-20.

[5]  Svelto, F.; Charbon, E.; Wilton, S.J.E, "Introduction to the special issue on the IEEE 2002 custom integrated circuits conference", University of Pavia.

[6]  James Nechvatal, Elaine Barker, Lawrence Bassham, William Burr, Morris Dworkin, James Foti, Edward Roback, "Report on the Development of the Advanced Encryption Standard (AES)", October 2, 2000.

[7]  Colin Percival, "Cache Missing for Fun and Profit", May 13, 2005.

[8]  Bruce Schneier, Doug Whiting (2000-04-07). "A Performance Comparison of the Five AES Finalists" (PDF/PostScript). Retrieved on 2006-08-13.

[9]  Niels Ferguson, Richard Schroeppel, Doug Whiting (2001). "A simple algebraic representation of Rijndael" (PDF/PostScript). Proceedings of Selected Areas in Cryptography, 2001, Lecture Notes in Computer Science: pp. 103–111, Springer-Verlag. Retrieved on 2006-10-06.

**Authors Biography**

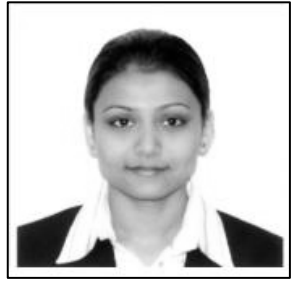

**Jalpa Bani** is a M.S. student of Computer Science at University of Bridgeport. She completed her under graduation in Computer Engineering from Saurashtra University, Gujarat, India. She has a deep urge to know more in the fields of Artificial Intelligence, Computer Networks, Database Management System and Mobile Computing. During her under graduation, she researched on Cyborg - an active area in applied Robotics. She continued her research quest by concentrating on security vulnerabilities in network and wireless communication protocols; 1024-bit+ encryption/decryption of data; and enhancing performance of mobile database query engine. In April 2008, she published an innovative paper - "A New Dynamic Cache Flushing (DCF) Algorithm for Preventing Cache Timing Attack" at IEEE Wireless Telecommunication Symposium (IEEE WTS 2008), Pomona, California. The paper presented a unique algorithm to prevent cache timing attack on Rijndael Algorithm. She also published a paper called "Adapting Anti-Plagiarism Tool into Coursework in Engineering Program," at American Society for Engineering Education (ASEE) at Austin, TX in June 2009. She achieved a "Best Student Poster" Honorable Mention in ASEE NE Conference at Bridgeport, CT in April 2009. She was honored with "School of Engineering Academic Achievement Award" at University of Bridgeport in May 2009.

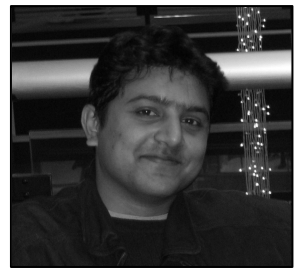

**Syed S. Rizvi** is a Ph.D. student of Computer Science and Engineering at University of Bridgeport. He received a B.S. in Computer Engineering from Sir Syed University of Engineering and Technology and an M.S. in Computer Engineering from Old Dominion University in 2001 and 2005, respectively. In the past, he has done research on bioinformatics projects where he investigated the use of Linux based cluster search engines for finding the desired proteins in input and outputs sequences from multiple databases. For last three year, his research focused primarily on the modeling and simulation of wide range parallel/distributed systems and the web based training applications. Syed Rizvi is the author of 68 scholarly publications in various areas. His current research focuses on the design, implementation and comparisons of algorithms in the areas of multiuser communications, multipath signals detection, multi-access interference estimation, computational complexity and combinatorial optimization of multiuser receivers, peer-to-peer networking, network security, and reconfigurable coprocessor and FPGA based architectures.